\documentstyle[aps,epsf,psfig,twocolumn]{revtex}

\tighten

\newcommand{\be}{\begin{equation}}
\newcommand{\ee}{\end{equation}}
\newcommand{\beq}{\begin{eqnarray}}
\newcommand{\eeq}{\end{eqnarray}}
\newcommand{\ho}{\rm\scriptscriptstyle}

\begin{document}

\draft
\twocolumn[\hsize\textwidth\columnwidth
\hsize\csname@twocolumnfalse\endcsname

\title{Information Tradeoff Relations for Finite-Strength
Quantum Measurements
\vbox to 0pt{\vss
                    \hbox to 0pt{\hskip-57pt\rm LA-UR-00-4403\hss}
                    \vskip 25pt}}
		    
\author{Christopher A. Fuchs$^\dagger$ and Kurt Jacobs\medskip}

\address{T-8, Theoretical Division, Los Alamos National
  Laboratory, Los Alamos, New Mexico 87545\\
  $^\dagger$Present Address: Bell Labs, Lucent Technologies,
  Murray Hill, New Jersey 07974}

\date{1 September 2000}

\maketitle

\begin{abstract}
In this paper we give a new way to quantify the folklore notion that
quantum measurements bring a disturbance to the system being
measured. We consider two observers who initially assign identical
mixed-state density operators to a two-state quantum system. The
question we address is to what extent one observer can, by
measurement, increase the purity of his density operator without
affecting the purity of the other observer's. If there were no
restrictions on the first observer's measurements, then he could
carry this out trivially by measuring the initial density operator's
eigenbasis. If, however, the allowed measurements are those of
finite strength---i.e., those measurements strictly within the
interior of the convex set of all measurements---then the issue
becomes significantly more complex.  We find that for a large class
of such measurements the first observer's purity increases the most
precisely when there is some loss of purity for the second observer.
More generally the tradeoff between the two purities, when it exists,
forms a monotonic relation. This tradeoff has potential application
to quantum state control and feedback.
\end{abstract}

 \vspace{5ex} ]

\section{Introduction}

Since the earliest days of quantum mechanics, a common idea
associated with the measurement process has been that it necessarily
disturbs or interferes with the system being observed. For instance
Bohr, in his reply to the Einstein-Podolsky-Rosen paper \cite{EPR},
writes that the quantum description ``may be characterized as a
rational utilization of all possibilities of unambiguous
interpretation \ldots\ compatible with the finite and uncontrollable
interaction between the objects and the measuring instruments''
\cite{Bohr35,CommentNonsense}. Or Pauli, on a much later occasion,
writes again, ``every act of observation is an interference, of
undeterminable extent, with the instruments of observation as well
as with the system observed, and interrupts the causal connection
between the phenomena preceding and succeeding it'' \cite{Pauli95}.
See Refs.~\cite{Jammer74,Beller99} for a more complete bibliographic
account of this issue.

Without question, it has also been apparent since the earliest days
of the theory that these proclamations are somewhat dubious. The
question is this: What is it that is being interfered with or
disturbed in a measurement? If there were a set of hidden variables
underneath the statistical predictions of quantum theory, then the
answer would be at hand: The act of measurement disturbs the hidden
variables. In the absence of a hidden-variable explanation
\cite{NoBohm}, however, this becomes a moot point.  Measuring {\it
x\/} cannot disturb {\it p\/} if {\it p\/} does not have an
independent existence before a measurement elicits its value
\cite{Mermin90}. In fact one has to wonder why the word
``measurement'' is used at all in this context: If there are no
free-standing values {\it x\/} and {\it p\/} to disturb, then surely
there are no values to measure either.

Eschewing metaphysical concerns, one might try to give a precise
sense to the idea that measurements cause disturbance by focusing
solely on the wavefunction itself.  For, the wavefunction appears to
be the simplest term in the theory that would even allow a precise
formulation of the question. One might say for instance, ``The word
measurement is a misnomer for our experimental interventions into
the course of nature \cite{Peres00,Fuchs00}. The unpredictable
wavefunction collapse is the quantitative signature of a disturbance
in quantum measurement. Since the state change is random, the
measurement causes an uncontrollable disturbance.'' But this
formulation too is not without problem. The quantum state resulting
from a measurement depends in a crucial way on the precise form of
the measurement interaction \cite{Braunstein88}.  In particular, if
there is only a single quantum state under scrutiny---as was the case
in the original Heisenberg uncertainty relation discussion
\cite{WheelerZurek}---or even an unknown state drawn from a fixed
orthogonal set \cite{Unruh78}, then a measurement interaction can
always be rigged for {\it any\/} observable so that, upon completion
of the process, the quantum state is returned to its initial value
\cite{Kraus83}.  It does not matter that the measurement outcome is
random and unpredictable: If the discussion is limited to a single
quantum state or an orthogonal set, then there need be no
disturbance in the sense of a {\it necessary\/} wavefunction change.

What appears to be needed is a situation where more than one quantum
state from within a nonorthogonal set arises naturally into the
considerations. Indeed, perhaps the first phenomenon to give a
precise meaning to the idea that information-gathering measurements
necessarily cause an accompanying disturbance is quantum cryptography
\cite{Wiesner83,Bennett84}. There it is essential that the systems
are known to be prepared in one or another quantum state drawn from
some fixed {\it nonorthogonal\/} set
\cite{Bennett92a,Bennett92b,Fuchs98}. These nonorthogonal states are
used to encode a potentially secret cryptographic key to be shared
between the sender and receiver. In this case, the information an
eavesdropper seeks is not about some nonexistent hidden variable like
$x$ or $p$, but instead about which quantum state was actually
prepared in each individual transmission. What is novel here is that
the encoding of the proposed key into nonorthogonal states forces the
information-gathering process to induce a disturbance to the overall
{\it set\/} of states. That is, the presence of an active
eavesdropper transforms the initial pure states into a set of mixed
states or, at the very least, into a set of pure states with larger
overlaps than before. This action ultimately boils down to a loss of
predictability for the sender over the outcomes of the receiver's
measurements and, so, is directly detectable by the receiver
revealing some of those outcomes for the sender's inspection. In
fact, there is a direct connection between the statistical
information gained by an eavesdropper and the consequent disturbance
she must induce to the quantum states in the process.  As the
information gathered goes up, the necessary disturbance also goes up
in a precisely formalizable way \cite{Fuchs96,Fuchs97,Bruss98}.

Note the two ingredients that appear in this formulation. First, the
information gathering or measurement is grounded with respect to one
observer (in this case, the eavesdropper), while the disturbance is
grounded with respect to another (here, the sender). In particular,
the disturbance is a disturbance to the sender's previous
information---this is measured by his diminished ability to predict
the outcomes of certain measurements the legitimate receiver might
perform. No hint of any variable intrinsic to the system is made use
of in this formulation.  In itself, this is already a rupture from
the founding fathers' description of disturbance in measurement. As
far as we can tell, all early literature on the subject refers the
discussion of disturbance exclusively to the system and the invasive
measuring device, not to the perspective of various observers
\cite{Jammer74}.

The second ingredient is another break with the founding fathers. One
must consider at least two possible non\-orthogonal preparations in
order for the formulation to have any meaning. This is because the
information gathering is not about some classically-defined
observable---i.e., about some unknown hidden variable or reality
intrinsic to the system---but is instead about which of the unknown
states the sender actually prepared.  The lesson is this: Forget
about the unknown preparation, and the random outcome of the quantum
measurement is information about nothing.  It is simply ``quantum
noise'' with no connection to any preexisting variable.

How crucial is this second ingredient, i.e., that there be at least
two nonorthogonal states within the set under consideration? We can
start to readdress its necessity by making a slight shift in the
account above. Divorcing the discussion from a cryptographic
protocol, one might say that the eavesdropper's goal is not so much
to uncover the identity of the unknown quantum state, but to sharpen
her predictability over the receiver's measurement outcomes.  In
fact, she would like to do this at the same time as disturbing the
sender's predictions as little as possible.  Changing the language
still further to the terminology of Ref.~\cite{Fuchs00}, the
eavesdropper's actions serve to sharpen her information about the
potential consequences of the receiver's further interventions upon
the system.  (Again, she would like to do this while minimally
diminishing the sender's previous information about those same
consequences.) In the cryptographic context, a byproduct of this
effort is that the eavesdropper ultimately comes to a more sound
prediction of the secret key. From the present point of view,
however, the importance of this change of language is that it leads
to an almost Bayesian perspective on the information--disturbance
problem \cite{Kyburg80}.

Within Bayesian probability theory, one of the overarching themes is
to identify the conditions under which a set of decision-making
agents can come to a common belief or probability assignment for
some specified random variable even though the agents' initial
beliefs may differ \cite{Bernardo94}.  One might similarly view the
process of quantum eavesdropping.  The sender and the eavesdropper
start off initially with differing quantum state assignments for a
single physical system.  In this case it so happens that the sender
can make sharper predictions than the eavesdropper about the
outcomes of the receiver's measurements.  The eavesdropper, not
satisfied with the situation, performs a measurement on the system
in an attempt to sharpen those predictions.  In particular, there is
an attempt to come into something of an agreement with the sender
but without revealing the outcomes of her measurements or, indeed,
her very presence.

It is at this point that a distinct {\it property\/} of the quantum
world makes itself known.  The eavesdropper's attempt to
surreptitiously come into alignment with the sender's predictability
is always shunted away from its goal.  This shunting of various
observer's predictability (and perhaps only this shunting
\cite{FuchsInfinity}) is the subtle manner in which the quantum
world is sensitive to our experimental interventions.

\begin{figure}
\centerline{\psfig{file=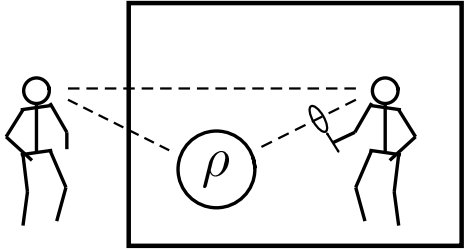,width=3.25in}}
\caption{\narrowtext Here two observers both ascribe a density matrix 
$\rho$ to a quantum system. The observer inside the box (Alice) makes 
a measurement on the system without telling the result to the observer 
outside (Bob). Alice wishes to obtain as much knowledge about the 
final state of the system as she can, while causing as little 
disturbance to Bob's state-of-knowledge as possible.}
\label{fig1}
\end{figure}

This motivates finally the following problem, which is the subject of
our paper. Suppose two players---let us call them Alice and Bob from
here out---come to agree about the way a quantum system will react
to any measurement.  In other words, by Gleason's theorem
\cite{Gleason57}, suppose they start with an identical density
operator assignment $\rho$ for the system. The case we are
interested in most is when $\rho$ is a mixed-state.  Under what
conditions can one player---Alice, say---surreptitiously increase
her knowledge of the system without forcing the other player's
knowledge to become less relevant? (See Fig.\ref{fig1})

To move toward making this question precise, imagine that a third
player will perform some measurement on the system in the future, but
neither Alice nor Bob know which it will be. Depending upon which
measurement is ultimately performed, Alice and Bob will have varying
degrees of predictability for its outcomes. For instance, consider
how their predictability fares with respect to various simple von
Neumann measurements. If the measurement happens to be the eigenbasis
of $\rho$, the Shannon entropy of the outcomes---which is a good
measure of predictability \cite{Ash65}---will be the minimal value it
can be \cite{Wehrl78}.  This turns out to be the von Neumann entropy
$S(\rho)=-{\rm tr}\,\rho\log\rho$. On the other hand, if the
measurement happens to be a ``mutually unbiased'' basis
\cite{Wootters89} to the eigenbasis, then all measurement outcomes
will be equally probable, and the outcome entropy will be $\log d$,
where $d$ is the dimension of the system's Hilbert space.

For the purpose at hand, we would like to capture in a single number
something about how much Alice and Bob can predict of the unknown
measurement. As a simple example, we might average the Shannon
entropy of the measurement outcomes over the unique unitarily
invariant measure (or ``uniform'' measure) on the space of von
Neumann measurements \cite{Wootters90,Jones91}.  This would represent
how well Alice and Bob will fare on average with respect to a
completely random von Neumann measurement.  Or, we might simply
consider the entropy of the best case scenario, i.e., the von Neumann
entropy of $\rho$ as above. Without getting specific---all will be
made precise later---we will generically call measures of this
flavor, measures of {\it purity}. The main intuition we want to
capture is that when $\rho=|\psi\rangle\langle\psi|$ is a pure state,
then Alice and Bob should generally have the most predictability over
the third party's measurements.  When $\rho$ is the ``completely
mixed state''---i.e., proportional to the identity operator,
$\rho=\frac{1}{d}I$---they should have the least.

The precise question we want to address is, can Alice secretly
increase the purity of her quantum state assignment at the same time
as leaving the other player's purity unscathed?  If she cannot, then
such a failure may hint at another interesting way to quantify a
quantum information--disturbance tradeoff.  The hallmark of this
formulation would be that it works even in the case where there is
only a {\it single\/} initial quantum state (albeit a mixed state),
while still capturing the shift in language we used to reformulate
the quantum eavesdropping process.

Unfortunately, the answer is trivial if we leave the question posed
in such a simplistic way.  For, we need only suppose that Alice
measures an eigenbasis $\Pi_b=|b\rangle\langle b|$ of $\rho$ to
derail the whole program. Upon finding some result $b$, Alice will
collapse her description of the system from the mixed state $\rho$
to the pure state \cite{Luders51}
\be
\rho_b = \frac{1}{p_b} \Pi_b \rho \Pi_b\;,
\ee
where $p_b={\rm tr}(\rho\Pi_b)$ is the probability of the particular
outcome. The upshot of this is to make Alice's final purity for the
system maximal, while as far as Bob is concerned the system's
density operator will not be affected at all. This is because, with
respect to Bob's state of knowledge, the quantum state evolves
simply to a mixture of Alice's states, i.e.,
\be
\rho\;\longrightarrow\;\sum_b p_b\rho_b = \rho\;,
\ee
and so his purity and indeed his quantum state assignment remain the
same.

The key to finding something interesting here is to ask what would
happen in the case where there is a well-justified restriction to
the class of measurements Alice can perform.  For instance, suppose
Alice has not yet reached the technologically advanced stage of
being able to perform a truly perfect von Neumann measurement. Maybe
she is using a finite temperature Stern-Gerlach device to perform a
spin measurement on an electron and, because of thermal noise, it
every now and then registers a spin to be down when it should have
registered a spin-up. To put this another way, instead of projecting
$\rho$ into the states $\Pi_0$ and $\Pi_1$ as Alice would like, the
Stern-Gerlach device projects $\rho$ according to a more general
L\"uder's rule for positive operator valued measures (POVMs)
\cite{Busch96,Busch98},
\be
\rho\;\longrightarrow\;\rho_b=\frac{1}{p_b}E_b^{1/2}\rho E_b^{1/2}
\ee
where in this case
\beq
E_0
&=&
\kappa\Pi_0+(1-\kappa)\Pi_1\
\label{jiggle}
\\
E_1
&=&
(1-\kappa)\Pi_0+\kappa\Pi_1\;,
\label{wiggle}
\eeq
and $p_b={\rm tr}(\rho E_b)$.  Similarly, the description of the
state change from Bob's perspective must be in accord with this, and
so is
\be
\rho\;\longrightarrow\;\tilde\rho=\sum_b p_b\rho_b\;.
\ee
When $\kappa$ is a number strictly between 0 and 1, we will call this
an instance of a {\it finite-strength\/} quantum measurement. (We use
this suggestive terminology because we imagine that Alice can never
really get to a perfect von Neumann measurement without the
expenditure of an infinite amount of effort.) What can be said in a
case like this?

Well, again, Alice will be able to generally increase the purity of
her state without causing any decrease to Bob's purity.  She does
this, as before, simply by choosing $\Pi_0$ and $\Pi_1$ to be
eigenprojectors of $\rho$. Then $E_0$ and $E_1$ commute with the
initial density operator, and it is straightforward to check that
$\tilde\rho=\rho$. However, we can now ask whether this is the
strategy that brings the greatest benefit to Alice. Might it be the
case that Alice can increase her purity even more on average if she
chooses $\Pi_0$ and $\Pi_1$ to be noncommuting with $\rho$? Moreover
if it does, what kind of havoc will that wreak on Bob's description
of the system? What we are imagining here in the imagery of the
Stern-Gerlach device is that though Alice may not be able to chill
her magnets to absolute zero, she can at least adjust their spatial
orientation at will.  Is this a freedom she should make use of?

Interestingly, it turns out that there is a tradeoff in the two final
purities. Whenever $\rho$ is nonpure (so that there is actually
something to be ``learned'') and $0<\kappa<1$ (so that the
measurement is of finite strength), Alice's final purity will be the
greatest on average precisely when Bob's purity has decreased the
most in turn. Moreover, varying through the class of measurements
that lead from the least average final purity to the most (with
respect to Alice), we find that Bob's purity goes down monotonically
as Alice's goes up. As we will show, this is an example of a more
general phenomenon where the measurement operators are not so
restricted as in Eqs.~(\ref{jiggle}) and (\ref{wiggle}): for a large
class of finite-strength quantum measurements, a nontrivial tradeoff
relation always exists.

The plan of the remainder of the paper is as follows.  In Section II,
we give a precise formulation of the problem in the widest setting,
including definitions of various measures of purity and also a
definition of the general notion of a finite strength quantum
measurement (without feedback).  In Section III, we work out an
analytic form for the changes of purity for both Alice and Bob under
the assumption of a particularly simple measure of purity {\it and\/}
the restriction that Alice's quantum measurements have only two
outcomes.  We then explore the various regimes of the convex set of
measurements and exhibit the general information tradeoff relation
where it exists.  We close in Section IV with a few concluding
remarks about the significance of this result. In Appendix A, we
prove that any efficient measurement (POVM) will increase Alice's
purity on average (for any measure of purity that is a convex
function of the density operator's eigenvalues)---this result is
essentially identical to one proven recently in
Ref.~\cite{Nielsen00}.  In Appendix B, for comparison with the main
result here, we consider a variation of the problem where we vary
over all measurements of a given finite strength instead of only
those on the unitary orbits of a given fiducial measurement.

\section{Formulation}

Our problem concerns two agents, Alice and Bob, who initially ascribe
a single density operator $\rho$ to a quantum system in which they
have some interest. The most important case for us is when $\rho$ is
a mixed state, i.e., ${\rm tr}\rho^2<1$.  For generality in the
formulation, let us assume that $\rho$ is a density operator over a
$d$-dimensional Hilbert space ${\cal H}_d$.  The detailed
considerations begin when Alice tries to surreptitiously increase her
``knowledge'' of the system---that is, to obtain a new density
operator that is closer to being a pure state than her initial
ascription.  The only way she can do this is by performing a quantum
measurement behind Bob's back.  To be as generous as we can be
without trivializing the problem, let us assume that Bob knows
everything of Alice's {\it plan\/}, even her precise measurement
interaction.  The only information barred from Bob is the precise
outcome of Alice's measurement.

The formalism for treating the most general kind of quantum
measurement is that of the positive operator valued measure, or POVM
\cite{Peres93}. In this formalism a measurement corresponds to a
sequence of operators on ${\cal H}_d$---denoted by ${\cal
E}=(E_b)_{b=1}^\infty$---with a {\it finite\/} number of
nonvanishing $E_b$, such that each element of the sequence is a
positive semi-definite operator, i.e.,
\be
\langle\psi|E_b|\psi\rangle\ge0\;,\quad\forall\,|\psi\rangle\;,
\ee
and, together, the elements form a resolution of the identity,
\be
\sum_b E_b = I\;.
\ee
The outcomes of the measurement are specified by the index $b$ and
occur with probabilities $p_b={\rm tr}\rho E_b$.

Upon finding an outcome $b$, the laws of quantum mechanics specify
that Alice's state can evolve into any other density operator of the
form \cite{Kraus83}:
\be
\rho\;\longrightarrow\; \rho_b = \frac{1}{p_b}\sum_i A_{bi}\rho
A_{bi}^\dagger\;,
\label{dogmeat}
\ee
where
\be
\sum_i A_{bi}^\dagger A_{bi}=E_b\;.
\label{hambone}
\ee
Since Bob knows nothing of Alice's outcome, as far as he is
concerned the state of the quantum system will evolve according to
\beq
\rho\;\longrightarrow\;\tilde\rho &=& \sum_b p_b \rho_b
\label{FidelityJohnson}
\\
&=& \sum_{b,i} A_{bi}\rho A_{bi}^\dagger\;.
\eeq

Note that the decomposition of each $E_b$ into the operators
$A_{bi}$ in Eqs.\ (\ref{dogmeat}) and (\ref{hambone}) depends
crucially upon the interaction Alice chooses for carrying out the
measurement $\cal E$. Whenever the range of the index $i$ is
restricted to a single value, we say that Alice's measurement is an
{\it efficient\/} one~\cite{WM}.

Efficient quantum measurements (with respect to a given $\cal E$)
correspond to holding on to as much information as possible in the
measurement process.  That is to say, such measurements do not break
quantum coherence more than is necessary for the given POVM. In the
subsequent development we will only consider efficient measurements
for just this reason.  Hence, in the language of equations, we will
only consider conditional state changes of the form
\be
\rho\;\longrightarrow\;\rho_b = \frac{1}{p_b} A_b\rho A_b^\dagger\;,
\label{Aharonov}
\ee
where $A_b^\dagger A_b=E_b$.

By the polar decomposition theorem for operators \cite{Schatten60},
we can always write
\be
A_b = U_b E_b^{1/2}\;,
\label{CupNoodle}
\ee
where $U_b$ is a unitary operator.  This decomposition can be endowed
with a physical meaning if one thinks of the measurement process as
allowing for a sort of {\it feedback\/} to the quantum system. The
raw measurement causes a ``collapse''
\be
\rho\;\longrightarrow\;\sigma_b=\frac{1}{p_b}E_b^{1/2}\rho E_b^{1/2}
\ee
in one's description of the system.  But then, conditioned upon the
outcome, one can think of the interaction as causing the system to
further unitarily evolve to
\be
\sigma_b\;\longrightarrow\;\rho_b=U_b\sigma_b U_b^\dagger\;.
\ee
This split, of course, is a conceptual one: it may or may not
correspond to the actual workings of the device carrying out the
measurement $\cal E$.  Nevertheless, it can be quite useful for
classifying different kinds of measurement interaction.

Efficient measurements {\it without\/} feedback hold a special place
in our considerations.  These are measurement interactions for which
all the $U_b=I$, so that Alice's state change is ultimately of the
simple form
\be
\rho\;\longrightarrow\;\rho_b=\frac{1}{p_b}E_b^{1/2}\rho E_b^{1/2}\;.
\ee
These hold a special place for us first and foremost because they
correspond to the ``rawest'' kind of measurement interaction allowed
for a given POVM\@.  Therefore, they are worthy of study in their own
right \cite{Busch98}\@. Secondly, though, there are other problems
for which they correspond to the least perturbing implementation of a
POVM\@. Namely, if one contemplates performing the measurement on a
system initially prepared in a completely random pure state, then
the mean input-output fidelity will be the greatest if the
measurement has no feedback \cite{Barnum98,Banaszek00}. Finally, it
stands to reason that if we can get a handle on the tradeoff between
information and disturbance for such a special case, we will be
better prepared for understanding the more general one of an
arbitrary efficient measurement. We will also be better prepared to
understand the precise role of feedback for controlling quantum
systems \cite{Doherty00}.

Our focus hereafter will be on efficient measurements without
feedback.  What is the {\it strength\/} of such a measurement?  This
issue is explored in Ref.~\cite{Doherty00}, where a more
refined notion of the concept is given a quantitative formulation.
For the purposes here, we will only need the rawest of distinctions:
finite vs.\ infinite measurement strength. An efficient measurement
$\cal E$ is said to be of finite strength as long as each
nonvanishing $E_b$ has support on the whole Hilbert space ${\cal
H}_d$---that is, as long as
\be
\mbox{rank}\,E_b=d\;, \quad\mbox{for all $b$ such that $E_b\ne0$}\;.
\ee
A measurement is of infinite strength any time one of the
nonvanishing $E_b$'s has rank strictly less than $d$.

The utility of this notion comes about from noting that the set of
all POVMs is a convex set.  This follows from the fact that one can
invent a notion of convex addition operation for POVMs:  Simply take
\cite{Fujiwara98}
\be
p{\cal E} + (1-p){\cal F}\equiv\big(p E_b + (1-p)
F_b\big)_{b=1}^\infty\;.
\ee
By a similar consideration, it is also true that the set of all POVMs
with a fixed number $n$ of nonvanishing elements $E_b$ is a convex
set. Thinking of this set as embedded in the space of length-$n$
sequences of all Hermitian operators, one has that the boundary of
such a set is given by precisely what we are calling the infinite
strength measurements (with $n$ outcomes). The finite strength
measurements lie strictly within the interior of the set. Making this
identification in terminology is an attempt to capture the idea that
an experimenter would need to expend an infinite amount of effort or
money to work his way out to the boundary. For, if he could get all
the way to the edge, there would be some preparations of the system
for which he could predict with {\it absolute\/} certainty that some
outcomes of the measurement would not occur. That strikes us as
beyond the power of mortals.  As technology advances, we can imagine
experimentalists getting ever closer to the boundary, but never
quite getting there.

Let us now start applying these distinctions of measurement to the
problem at hand---namely, to that of an Alice trying to
surreptitiously increase her ``knowledge'' of a system while
affecting Bob's ``knowledge'' of it as little as possible.  How shall
we quantify ``knowledge'' in this context?  There are at least three
canonical ways.

The first has to do with the von Neumann entropy of a density
operator $\rho$:
\be
S(\rho) = - {\rm tr}\,\rho\log\rho
=-\sum_{k=1}^d\lambda_k\log\lambda_k\;,
\label{Lahti}
\ee
where the $\lambda_k$ signify the eigenvalues of $\rho$.  (We
evaluate all logarithms in base 2 so that information is measured in
bits, rather than nats or hartleys \cite{Gallager68}.  Also, we use
the convention that $\lambda\log\lambda=0$ whenever $\lambda=0$ so
that $S(\rho)$ is always well defined.)

The intuitive meaning of the von Neumann entropy can be found by
first thinking about the Shannon entropy. Consider any von Neumann
measurement $\cal P$ consisting of $d$ one-dimensional orthogonal
projectors $\Pi_i$. The Shannon entropy for the outcomes of this
measurement is given by
\be
H({\cal P})=-\sum_{i=1}^d \big({\rm tr}\rho\Pi_i\big)\log\big({\rm
tr}\rho\Pi_i\big)\;.
\ee
This number is bounded between 0 and $\log d$, and there are several
reasons to think of it as a good measure of {\it impredictability\/}
over the outcomes of a measurement $\cal P$. Perhaps the most
important of these is that it quantifies the number of {\it
yes-no\/} questions one can expect to ask per measurement, if one's
only means to ascertain the measurement outcome is from a colleague
who knows the actual result \cite{Ash65}. Under this quantification,
the lower the Shannon entropy, the more predictable a measurement's
outcomes.

A natural question to ask is:  With respect to a given density
operator $\rho$, which measurement $\cal P$ will give the most
predictability over its outcomes? As it turns out, the answer is any
$\cal P$ that forms a set of eigenprojectors for $\rho$
\cite{Wehrl78}. When this obtains, the Shannon entropy of the
measurement outcomes reduces to simply the von Neumann entropy of
the density operator. The von Neumann entropy, then, signifies the
amount of impredictability one achieves by way of a standard
measurement in a best case scenario.  Indeed, true to one's
intuition, one has the most knowledge by this account when $\rho$ is
a pure state---for then $S(\rho)=0$.  Alternatively, one has the
least knowledge when $\rho$ is proportional to the identity
operator---for then any measurement $\cal P$ will have outcomes that
are all equally likely.

The best case scenario for predictability, however, is a very limited
case, and not so very informative about the density operator as a
whole.  Since the density operator contains, in principle, all that
can be said about every possible measurement \cite{Gleason57}, it
seems a shame to throw away the vast part of that information in our
considerations.

This issue leads to our next quantification of ``knowledge'' of a
quantum system. For this, we again rely on the Shannon information
as our basic notion of predictability.  The difference is we
evaluate it with respect to a ``typical'' measurement rather than
the best possible one.  However with this, a new question arises:
Typical with respect to what?  The notion of typical is only defined
with respect to a given {\it measure\/} on the the set of
measurements.

Luckily, there is a fairly canonical answer. There is a unique
measure $d\Omega_\Pi$ on the space of one-dimensional projectors
that is invariant with respect to all unitary operations.  That in
turn naturally induces a canonical measure $d\Omega_{\cal P}$ on the
space of von Neumann measurements $\cal P$
\cite{Wootters90,Jones91}. Using this measure gives rise to the
following quantity
\begin{eqnarray}
\overline{H}(\rho)&=&\int H(\Pi)\,d\Omega_{\cal P}
\\
&=& -d \int \big({\rm tr}\rho\Pi\big)\log\big({\rm
tr}\rho\Pi\big)\,d\Omega_\Pi\;,
\end{eqnarray}
which is intimately connected to the so-called quantum
``subentropy'' of Ref.~\cite{Jozsa94}.  Interestingly, this mean
entropy can be evaluated explicitly in terms of the eigenvalues of
$\rho$ and takes on the expression
\be
\overline{H}(\rho)=\frac{1}{\ln2}\left(\frac{1}{2}+
\frac{1}{3}+\cdots+\frac{1}{d}\right)+ Q(\rho)
\ee
where the subentropy $Q(\rho)$ is defined by
\be
Q(\rho)=-\sum_{k=1}^d\! \left(\prod_{i\ne
k}\frac{\lambda_k}{\lambda_k-\lambda_i}\right)\!\lambda_k\log\lambda_k\;
.
\label{Mittelstaedt}
\ee
In the case where $\rho$ has degenerate eigenvalues,
$\lambda_l=\lambda_m$ for $l\ne m$, one need only reset them to
$\lambda_l+\epsilon$ and $\lambda_m-\epsilon$ and consider the limit
as $\epsilon\rightarrow0$.  The limit is convergent and hence
$Q(\rho)$ is finite for all $\rho$.  With this, one can also see
that for a pure state $\rho$, $Q(\rho)$ vanishes. Furthermore, since
$\overline{H}(\rho)$ is bounded above by $\log d$, we know that
\be
0\le Q(\rho)\le\log d - \frac{1}{\ln2}\!\left(\frac{1}{2}+
\cdots+\frac{1}{d}\right)\le\frac{1-\gamma}{\ln 2}\;,
\ee
where $\gamma$ is Euler's constant.  This means that for any $\rho$,
$Q(\rho)$ never exceeds 0.60995 bits.

The interpretation of this result is the following.  Even when one
has the maximal knowledge about a system one can have under the laws
of quantum mechanics---i.e., when one has a pure state---one can
predict almost nothing about the outcome of a typical measurement
\cite{Caves96}.  In the limit of large $d$, the outcome entropy for a
typical measurement is just a little over a half bit away from its
maximal value.  Having a mixed state for a system, reduces one's
predictability even further, but indeed not by that much: The small
deviation is captured by the function in Eq.~(\ref{Mittelstaedt}),
which becomes a quantification of ``knowledge'' in its own right.

The two quantifications of knowledge about a quantum system given by
Eqs.~(\ref{Lahti}) and (\ref{Mittelstaedt}) are without doubt two of
the most well-motivated such quantities.  However, because of their
particular mathematical structures (involving logarithms and ratios
of eigenvalues, etc.), they are often difficult to work with.  It is
therefore useful to consider quantities $F(\rho)$ that may not have
the strictest of interpretations in terms of ``knowledge'' or
``information,'' but nevertheless carry some of the properties
essential to the explorations we would like to make.  The two
properties that appear to be the most important for us is that a
function $F$ from density operators to real numbers be (1) unitarily
invariant so that it only depends upon the eigenvalues of the
density operator, and (2) concave in its argument. That is, one
should have
\be
F\big(p\rho_0+(1-p)\rho_1\big)\ge p F(\rho_0) +(1-p)F(\rho_1)\;,
\label{Busch}
\ee
for each pair of density operators $\rho_0$ and $\rho_1$, and each
real number $p$ in the range $[0,1]$.

A common way of simplifying problems to do with the Shannon entropy
is to consider instead a function that is merely quadratic in the
probabilities \cite{Jaynes57,Aczel75}. In quantum mechanical terms,
this translates to a function we shall call the {\it impurity\/} of
a quantum state:
\be
P(\rho)= 1-{\rm tr}(\rho^2) = 1-\sum_{k=1}^d \lambda_k^2 \;.
\label{ChastityGoodhope}
\ee
This function, of course, has our two desired properties
\cite{Wehrl78}.  Moreover, it attains its minimum value of 0 when
$\rho$ is a pure state (just as $S$ and $Q$ do), and it
attains its maximum value of $\frac{d-1}{d}$ when $\rho$ is the
completely mixed state.

What makes unitarily invariant functions like the $F(\rho)$ in
Eq.~(\ref{Busch}) special is that one can prove an interesting
theorem for them in the measurement context. Consider any efficient
measurement of a POVM ${\cal E}=\{E_b \}$. Upon finding an outcome
$b$, the observer will update his quantum state for the system from
the original $\rho$ to some $\rho_b$ of the form in
Eq.~(\ref{Aharonov}).  What does this say about his expected change
of knowledge?  Well, one can prove that, whatever $F$ is,
\be
F(\rho)\ge\sum_b p_b F(\rho_b)\;.
\label{Blokhintsev}
\ee
In particular, it follows that
\beq
S(\rho) &\ge& \sum_b p_b S(\rho_b)\;,
\\
Q(\rho) &\ge& \sum_b p_b Q(\rho_b)\;,
\\
P(\rho) &\ge& \sum_b p_b P(\rho_b)\;.
\eeq
These statements---and in fact a stronger statement to do with a
majorization relation between the eigenvalues of $\rho$ and those of
the $\rho_b$---will be proven in Appendix A.  (See also Nielsen
\cite{Nielsen00} for an earlier proof of this result.)

The fact that Eq.~(\ref{Blokhintsev}) holds for all concave functions 
$F$ expresses what is meant by the phrase
``the observer {\it learns\/} something from a quantum measurement''
\cite{Nielsen00b}. Note in particular that it need not necessarily
be the case that the purity, etc., be nondecreasing in any {\it
individual\/} trial of a measurement. A simple counterexample
suffices for illustration. Take,
\be
 \rho=\frac{1}{3}\!\pmatrix{
  1 & 0 \cr
  0 & 2}
\ee
and consider a two-outcome efficient measurement without feedback
${\cal E}=(E,I-E)$ where
\be
E=\frac{1}{3}\!\pmatrix{
  2 & 0 \cr
  0 & 1}\;.
\ee
Note that if outcome $E$ occurs, the updated density operator for
the system will be the completely mixed state
\be
\rho_{\ho E}=\frac{1}{2}\!\pmatrix{
  1 & 0 \cr
  0 & 1}
\ee
which is certainly less pure than the initial state.  Thus, one can
only expect one's ``knowledge'' to increase {\em on average} during a 
measurement.

Going back to our target scenario with Alice and Bob, one can see
that this result insures that Alice comes away on average with more
information than she started with.  Moreover, this holds
independently of the particular way in which we choose to quantify
her ``information.'' To make some notation, this means that the
quantities
\be
\Delta^{\!\ho F}_{\ho in}\equiv F(\rho)-\sum_b p_b F(\rho_b)
\ee
will all be nonnegative for any efficient measurement.  The
subscript on $\Delta^{\!\ho F}_{\ho in}$ denotes that this refers to
the change of knowledge from the ``inside'' point of view of the
measurer.

An almost dual result is that from Bob's point of view---the outside
point of view---whenever $\cal E$ is not only an efficient
measurement, but also a measurement {\it without feedback}, his
information can never increase from Alice's actions.  That is to say,
using notation from Eq.~(\ref{FidelityJohnson}), the quantity
\be
\Delta^{\!\ho F}_{\ho out}\equiv F(\tilde\rho)-F(\rho)
\ee
is nonnegative for all concave unitarily invariant functions $F$
\cite{Ando89}.  Again, the subscript in $\Delta^{\!\ho F}_{\ho out}$
makes explicit that we are referring to a change of knowledge from
the outside point of view. (The interested reader can find a proof
that $\Delta^{\!\ho F}_{\ho out}\ge0$ in Ref.~\cite{Ando89}.)

We must emphasize that this result is {\it almost\/} dual to
Eq.~(\ref{Blokhintsev}): for it certainly depends upon the
assumption that the measurement is without feedback.  Let us show
this by way of a quick counterexample.  Take $\cal E$ to be a
complete set of orthogonal projectors $E_b=|b\rangle\langle b|$,
$b=1,\ldots,d$.  One possible measurement with feedback that is
consistent with this POVM is given by taking $A_b=|\psi\rangle\langle
b|$ for some fixed unit vector $|\psi\rangle$.  Clearly $A_b^\dagger
A_b=E_b$ as required.  However,
\be
\tilde\rho=\sum_b A_b\rho A_b^\dagger=|\psi\rangle\langle\psi|\;,
\ee
completely independently of what the initial state $\rho$ is.  So it
can certainly be the case that $F(\tilde\rho)\le F(\rho)$ if one
allows feedback into the picture.

The conclusion to draw is that we are right on track in considering
the quantities $\Delta^{\!\ho F}_{\ho in}$ and $\Delta^{\!\ho
F}_{\ho out}$ in the context of measurements without feedback:  in a
sense, they are compensatory of each other.  What we would like to
do now is sharpen that idea.  For, just because Alice's knowledge of
the system can only increase through her measurements and Bob's can
only decrease, it does not follow that there is necessarily a
monotonic relation between these adjustments.

Here is how we will tackle the problem explicitly.  As has been the
case since the beginning, we imagine the initial state of knowledge
for Alice and Bob fixed to be some density operator
$\rho$.  Now, however, we introduce a {\it fiducial\/}
quantum measurement ${\cal M}=(M_b)$ that will also be fixed
throughout our considerations.  The freedom we give Alice is that
she may perform any measurement without feedback that is unitarily
equivalent to $\cal M$.  That is to say, we shall consider
measurement operators for Alice that are necessarily of the form
\be
E_b=U M_b U^\dagger\;,
\ee
where $U$ is any unitary operation.  Each different $U$ defines a
consequent change in both Alice and Bob's total information which we
denote by $\Delta^{\!\ho F}_{\ho in}(U)$ and $\Delta^{\!\ho F}_{\ho
out}(U)$, respectively.  (This notation makes no reference to $\rho$
and $\cal M$ because they are fixed background information for the
problem.)  What we would like to know is:  Under what conditions is
there a nontrivial monotone relation between $\Delta^{\!\ho F}_{\ho
in}(U)$ and $\Delta^{\!\ho F}_{\ho out}(U)$ as we vary $U$?  In the
cases where such a monotone relation exists, that will be the
tradeoff we have been seeking.

This completes the formulation of our problem. Unfortunately, as
opposed to the formulation, we have not settled the issue of a
tradeoff relation in complete generality.  Study of the
two-dimensional Hilbert-space case, however, already turns out to be
of significant interest. In the next Section, we report a careful
study of the case where $d=2$ and $\cal M$ contains two outcome
operators $M_0$ and $M_1$. Even in this restricted class, there is a
large regime of measurements with a nontrivial information tradeoff
relation.

\section{The 2-D Two-Outcome Problem}

In this section, we assume explicitly that $d=2$, so that Alice and
Bob's information is about a single qubit.  The canonical
measurement $\cal M$ that sets Alice's standard is taken to consist
of only two elements $M_0$ and $M_1$, but is otherwise completely
general. Alice now has the freedom to choose any unitary operation
$U$, and consequently perform any measurement $\cal E$ consisting of
elements $E_0=UM_0U^\dagger$ and $E_1=UM_1U^\dagger$.  The question
we should like to address is how $\Delta^{\!\ho F}_{\ho in}(U)$ and
$\Delta^{\!\ho F}_{\ho out}(U)$ change with respect to each other as
a function of $U$.

Note that because there are only two outcomes to the measurement,
$E_0$ and $E_1$ must commute.  There is therefore only one
diagonalizing basis required in specifying this measurement.  Let us
relabel the measurement to make that more explicit:  We shall simply
denote the two outcomes by $E$ and $I-E$.  With our previous
definitions, this measurement is of finite strength when neither $E$
nor $I-E$ is a rank-1 operator.

We have performed extensive numerical work that shows the following
when $\rho$ is impure and $\cal E$ is of finite strength. For the
three concave functions $S(\rho)$, $Q(\rho)$, and $P(\rho)$
considered in the previous section, there are significant regions in
POVM space where $\Delta^{\!\ho F}_{\ho in}(U)$ achieves its maximum
value precisely when $\Delta^{\!\ho F}_{\ho out}(U)$ is nonminimal.
That is to say, Alice cannot learn the most unless she also disturbs
Bob's information in the process. In this situation, the optimal
measurement operator $E$ does not commute with $\rho$.
Alternatively, when $E$ commutes with $\rho$, the difference
$\Delta^{\!\ho F}_{\ho out}(U)$ achieves its minimum value, namely
$0$---so that Bob's information is not disturbed at all---but then
$\Delta^{\!\ho F}_{\ho in}(U)$ achieves its minimum value too---so
that Alice has learned the least amount possible.  In general, the
functional relationship $\Delta^{\!\ho F}_{\ho
out}\!\big(\Delta^{\!\ho F}_{\ho in}\big)$ is a monotonic one as
$\Delta^{\!\ho F}_{\ho in}$ ranges from its minimum to its maximum
value.  In those regions of POVM space where there is no nontrivial
tradeoff relation, the curve for $\Delta^{\!\ho F}_{\ho
out}\!\big(\Delta^{\!\ho F}_{\ho in}\big)$ is simply flat.

What we shall do herein is focus on quantifying the tradeoff
explicitly for the case in which ``knowledge'' is identified with the
impurity function $P(\rho)$ of Eq.~(\ref{ChastityGoodhope}). In this
case, all calculations can be done analytically and one can get a
feel for the exact form of things.  (In the other cases of $F=S$ or
$F=Q$, things are not terribly worse, but because the binary Shannon
entropy function cannot be inverted analytically, there is no way to
get an analytic expression for the function $\Delta^{\!\ho F}_{\ho
out}\!\big(\Delta^{\!\ho F}_{\ho in}\big)$\@.)  With this
restriction, we will hereafter drop the superscript $F$ from our
notation and write simply $\Delta_{\ho out}$ and $\Delta_{\ho in}$
for the ``information'' changes we are considering.

Let us start the calculations straight away.  From the inside point
of view of Alice, the two possible state changes are of the form
\be
\rho\;\longrightarrow\;\rho_{\ho E}=\frac{1}{{\rm tr}\rho
E}\,\sqrt{E}\,\rho\,\sqrt{E}\;,
\ee
and
\be
\rho\;\longrightarrow\;\rho_{\ho \neg E}=\frac{1}{1-{\rm tr}\rho
E}\,\sqrt{I-E}\,\rho\,\sqrt{I-E}\;.
\ee
>From the outside point of view of Bob, it is simply
\be
\rho\;\longrightarrow\;\tilde\rho=\sqrt{E}\,\rho\,\sqrt{E}+
\sqrt{I-E}\,\rho\,\sqrt{I-E}\;.
\ee
Keeping in mind that $\sqrt{E}$ and $\sqrt{I-E}$ commute, a little
algebra yields that
\beq
\Delta_{\ho out} &=& {\rm tr}\,\rho^2-{\rm tr}\,\tilde{\rho}^2
\\
&=& 2\Big[\,{\rm tr}\,\rho^2E - {\rm tr}\,\rho E\rho E  \nonumber
\\
&& \qquad -\, {\rm tr}\!
\left(\rho\,\sqrt{E(I-E)}\,\rho\,\sqrt{E(I-E)}\,\right)\Big]\;.
\label{Hambone}
\eeq
Similarly,
\beq
\Delta_{\ho in} &=& {\rm tr}\,\rho E\,{\rm tr}\,\rho_{\ho E}^2\, +\,
{\rm tr}\,\rho(I-E)\,{\rm tr}\,\rho_{\ho \neg E}^2\, -\, {\rm
tr}\,\rho^2
\\
&=& \frac{1}{{\rm tr}\,\rho E\,(1-{\rm tr}\,\rho E)}\Big[\,{\rm
tr}\,\rho E\rho E + {\rm tr}\,\rho^2\,{\rm tr}\,\rho E  \nonumber
\\
&& \qquad\qquad\qquad\qquad -\, 2\, {\rm tr}\,\rho E\, {\rm
tr}\,\rho^2 E\,\Big]-\, {\rm tr}\,\rho^2
\\
&=& \frac{1}{{\rm tr}\,\rho E\,(1-{\rm tr}\,\rho E)}\Big[\, {\rm
tr}\,\rho E\rho E - 2\, {\rm tr}\,\rho E\, {\rm tr}\,\rho^2 E
\nonumber
\\
&& \qquad\qquad\qquad\qquad + {\rm tr}\,\rho^2\,\big({\rm tr}\,\rho
E\big)^2\Big]\;.
\label{Skeezix}
\eeq
Note immediately that if $E$ and $\rho$ commute, then $\Delta_{\ho
out}$ vanishes as one would expect.

Since we are dealing with a two-dimensional Hilbert space, it is
most convenient at this point to switch to a kind of Bloch-sphere
notation for all operators.  Then we may write,
\be
\rho = \frac{1}{2}\big(I+\vec{a}\cdot\vec{\sigma}\big)\;,
\ee
where $\vec{a}=(a_x, a_y, a_z)$ is some vector of real numbers with
modulus $a\le1$ and $\vec{\sigma}$ is the vector of Pauli operators.
Similarly, if $\alpha={\rm tr}E$, then the operator
\be
B=\frac{1}{\alpha}E
\label{Mongol}
\ee
is a density operator, and we may write
\be
B = \frac{1}{2}\big(I+\vec{b}\cdot\vec{\sigma}\big)\;,
\label{Mania}
\ee
where $\vec{b}$ also has a length $b$ no greater than unity.  In
this notation, $E$ and $\rho$ commute if and only if $\vec{b}$ and
$\vec{a}$ lie within the same ray.

Since $0\le E\le I$, we must have $0\le\alpha\le 2$. Moreover, we
must insure that the larger eigenvalue of $E$ is no greater than
unity. Using the fact that the eigenvalues of $E$ in Bloch-sphere
notation are given by $\frac{1}{2}\alpha(1\pm b)$, it follows that
we must require
\be
\alpha\le\frac{2}{1+b}\;.
\label{Mojo}
\ee
One can see $\cal E$ becomes an infinite strength measurement
whenever $b=1$ and $\alpha$ is any value, or whenever $b<1$ but
$\alpha=2/(1+b)$.  The parameter $\alpha$ to some extent captures
the amount of symmetry between the two measurement operators $E$ and
$I-E$.  It is therefore natural to call the case where $\alpha=1$
the {\it symmetric case}.

With the notations of Eqs.~(\ref{Mongol}) and (\ref{Mania}) it
becomes a tractable task to calculate the various operators in
Eqs.~(\ref{Hambone}) and (\ref{Skeezix}). Using the law of
multiplication for Pauli matrices, i.e.,
\be
(\vec{m}\cdot\vec{\sigma})(\vec{n}\cdot\vec{\sigma})=
(\vec{m}\cdot\vec{n})I+ i\vec{\sigma}\cdot(\vec{m}\times\vec{n})\;,
\ee
one finds fairly easily that:
\beq
{\rm tr}\rho^2 &=& \frac{1}{2}(1+a^2)
\\
{\rm tr}\rho E &=& \frac{\alpha}{2}(1+abz)
\\
{\rm tr}\rho^2E &=& \frac{\alpha}{4}\big(1+a^2+2abz\big)
\\
{\rm tr}\rho E\rho E &=& \frac{\alpha^2}{8}\big(1+a^2+b^2
+a^2b^2(2z^2-1)+ 4abz\big)
\eeq
where $z=\cos\theta$, and $\theta$ is the angle between the vectors
$\vec{a}$ and $\vec{b}$.

The only really daunting term that we must calculate is the quantity
\be
{\rm tr}\! \left(\rho\sqrt{E(I-E)}\rho\sqrt{E(I-E)}\right)\;.
\ee
To make some headway, let
\be
G \equiv E(I-E) = g_0 I + \vec{g}\cdot\vec{\sigma}
\label{miasma}
\ee
where
\beq
g_0 &=& \frac{1}{4}\alpha\big(2-\alpha-\alpha b^2\big)\,
\\
\vec{g} &=& \frac{1}{2}\alpha(1-\alpha)\vec{b}\;.
\eeq
We need to find an $r_0$ and $\vec{r}$ such that
\be
\sqrt{G}=r_0 I + \vec{r}\cdot\sigma\;.
\ee
The method for this is simple:  We just need calculate
$G=\sqrt{G}\sqrt{G}$ and set the resultant equal to
Eq.~(\ref{miasma}).  Carrying this procedure to its conclusion, we
arrive at the following identifications:
\be
r_0^2 = \frac{\alpha}{8}\!\left[\,2-\alpha-\alpha b^2 +
\sqrt{(1-b^2)\big(4-4\alpha+(1-b^2)\alpha^2\big)}\,\right]
\ee
\be
\vec{r}=\frac{1}{4r_0}\,\alpha(1-\alpha)\vec{b}\;.
\ee
With this, we can finally calculate
\be
{\rm tr}\,\sqrt{G}\rho\sqrt{G}\rho =
\frac{1}{2}\!\Big[1+a^2+c^2+2(\vec{a}\cdot\vec{c})^2-a^2c^2+
4\vec{a}\cdot\vec{c}\,\Big],
\ee
where the vector $\vec{c}$ and its magnitude $c$ are defined by
\be
\vec{c}=\frac{\vec{r}}{r_0}\;.
\ee

Putting all these ingredients together, we finally arrive at our
sought-after expressions:
\be
\Delta_{\ho in} = \frac{\alpha b^2 (1-a^2)(1-a^2
z^2)}{2(1+abz)(2-\alpha-\alpha a b z)}\;,
\label{JoelUnderwood}
\ee
and
\be
\Delta_{\ho out} = \frac{1}{2}\!\left(\frac{\alpha a
b}{2r_0}\right)^{\! 2}\!
\Big[(1-\alpha)^2+4r_0^2\Big]\big(1-z^2\big)\;.
\label{DaleUnderwood}
\ee
These two equations contain everything needed for a complete analysis
of the information tradeoff question.  Let us first see how this
plays out for the simple case described in Eqs.~(\ref{jiggle}) and
(\ref{wiggle}) of the Introduction.

\subsection{The Symmetric Case}

In this case the measurement operators $M_0$ and $M_1$ take the form
\beq
M_0 &=& \kappa\Pi_0+(1-\kappa)\Pi_1\
\label{diggle}
\\
M_1 &=& (1-\kappa)\Pi_0+\kappa\Pi_1\;,
\label{piggle}
\eeq
where $0<\kappa<1$, and $\Pi_0$ and $\Pi_1$ are the projectors onto
some orthonormal basis.  Measurement operators of this form come up
quite naturally in the theory of continuous quantum measurements
\cite{Doherty00}. In our Bloch sphere notation of Eqs.~(\ref{Mongol})
and (\ref{Mania}), this case corresponds to taking $\alpha=1$ and
$b=2\kappa-1$.

Plugging $\alpha=1$ into Eqs.~(\ref{JoelUnderwood}) and
(\ref{DaleUnderwood}), we find the significantly simpler expressions
\be
\Delta_{\ho in} = \frac{1}{2}b^2\big(1-a^2\big)\frac{1-a^2 z^2}{1-a^2
b^2 z^2}
\label{Beebop}
\ee
and
\be
\Delta_{\ho out} = \frac{1}{2}b^2 a^2 \big(1- z^2\big)\;.
\label{Scat}
\ee
Clearly, $\Delta_{\ho out}$ is minimized when $z=1$ or $-1$ (so that
$E$ commutes with $\rho$) as we have noted before.  Moreover
$\Delta_{\ho out}$ is maximized when $z=0$---that is to say, when the
operator $E$ is diagonal in a basis complementary or mutually
unbiased to the diagonal of $\rho$.  On the other hand, since $b\le
1$, $\Delta_{\ho in}$ is a strictly decreasing function in $z^2$.
This gives $\Delta_{\ho in}$ the same qualitative behavior as
$\Delta_{\ho out}$ and ultimately leads precisely to our tradeoff
relation:  Eliminating $z^2$ from Eqs.~(\ref{Beebop}) and
(\ref{Scat}), we obtain
\be
\Delta_{\ho out} = \frac{2(1-a^2b^2)\Delta_{\ho
in}-b^2(1-a^2)^2}{2\big(1-a^2-2\Delta_{\ho in}\big)}\;.
\label{NoDiscount}
\ee

This example is something of an extreme for the phenomena we have
been hoping for.  As long as the fiducial measurement $\cal M$ is of
finite strength (i.e., $b\ne1$), Eq.~(\ref{NoDiscount}) traces out a
nontrivial monotone curve as we go from $\Delta_{\ho in}^{\ho \!
min}$ to $\Delta_{\ho in}^{\ho \! max}$. But more than this,
$\Delta_{\ho in}$ is maximized at precisely the same value of $z$
for which $\Delta_{\ho out}$ is also maximized.  In common language,
this means that if Alice wishes to gather the most information, she
must reciprocally cause Bob to loose the most information that her
class of measurements will allow.  The only means for Alice to
lessen the sting of this effect is to develop her technology so that
the limit $b\rightarrow 1$ can be approached asymptotically.

This behavior, at first sight, appears to be quite deep.  It helps
lend credence to the idea that measurements without feedback are
always somewhat destructive by their nature---that is, as long as
one's aim is to increase one's information as much as possible under
the constraint of having less than ``infinitely powerful''
measurement devices.

Interestingly, however, this type of behavior is not completely
generic.  There are some fiducial measurements $\cal M$ of finite
strength for which the tradeoff effect disappears.  To see this, we
must turn back to our base equations Eqs.~(\ref{JoelUnderwood}) and
(\ref{DaleUnderwood}).

\subsection{The General Case}

Let us now assume strictly that none of the variables $a$, $b$, or
$\alpha$ happen to equal unity. Then, as in the symmetric case, the
quantity $\Delta_{\ho out}$ is clearly minimized when $z^2=1$.
Similarly, the disturbance to Bob's knowledge is largest when $z=0$,
so that $E$ is diagonal in a basis mutually unbiased with respect to
the diagonal of $\rho$.

The analysis of the general $\Delta_{\ho in}$ is significantly more
difficult. One can show that the quantity is minimized at $z^2=1$,
but whether that occurs at $z=1$ or $z=-1$ now depends upon the size
of $\alpha$.  The way to see this is by checking that $\Delta_{\ho
in}$ is concave as a function of $z$: The calculation is tedious,
but it can be done analytically.  The point where the curve changes
from a positive slope to a negative slope, i.e., where the function
attains its maximum, is given by
\be
z=z_0=\frac{1}{\alpha(1-\alpha)a b}\Big[4 r_0^2 -
\alpha\big(2-\alpha-\alpha b^2\big)\Big]\;.
\label{Zombify}
\ee

This expression is quite revealing.  For a fixed value of $b\ne0$,
one can check for those values of $\alpha$ that force $z_0=1$ or
$z_0=-1$.  These are
\be
\alpha|_{z_0=1}= \frac{b(1+a^2)+2a}{b(1+a^2)+a(1+b^2)}
\ee
and
\be
\alpha|_{z_0=-1}= \frac{b(1+a^2)-2a}{b(1+a^2)-a(1+b^2)}\;.
\ee
This means that for $\alpha$ in the ranges
\be
0\, \le \,\alpha\, \le\,\max\!
\left\{0,\,\frac{b(1+a^2)-2a}{b(1+a^2)-a(1+b^2)}\right\}
\label{Django}
\ee
and
\be
\frac{b(1+a^2)-2a}{b(1+a^2)-a(1+b^2)}\, \le \,\alpha\, \le\,
\frac{2}{1+b}
\label{Reinhardt}
\ee

\begin{figure}
\centerline{\psfig{file=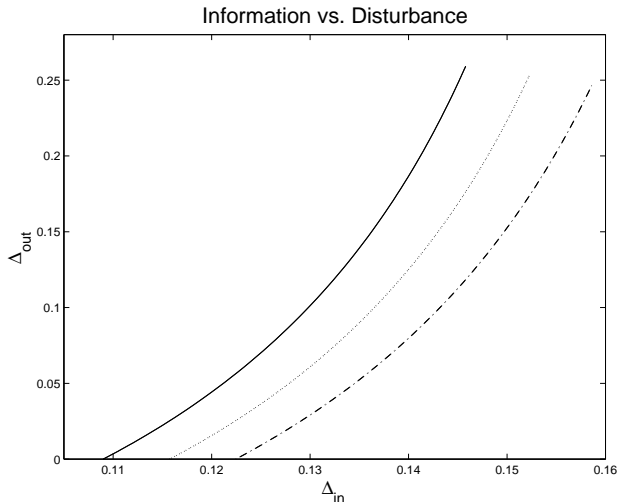,width=3.25in}}
\caption{\narrowtext The tradeoff between information, 
$\Delta_{\ho in}$, and disturbance, $\Delta_{\ho out}$, is
plotted here for \mbox{$\alpha=1$}, \mbox{$b = 0.9$} and
three values of \mbox{$a$}. Solid line: \mbox{$a=0.8$}, 
dotted line: \mbox{$a=0.79$} and dot-dash line: 
\mbox{$a=0.78$} .}
\label{fig2}
\end{figure}

$\!\!\!\!\!\! \Delta_{\ho in}$ will always be maximized by 
choosing $z=1$ or
$z=-1$.  However, for $\alpha$ outside of either of those ranges,
there will always be a nontrivial tradeoff relation:  When Alice's
information gain $\Delta_{\ho in}$ is maximized, Bob's information
loss $\Delta_{\ho out}$ will be strictly greater than its minimal
value.

\begin{figure}
\centerline{\psfig{file=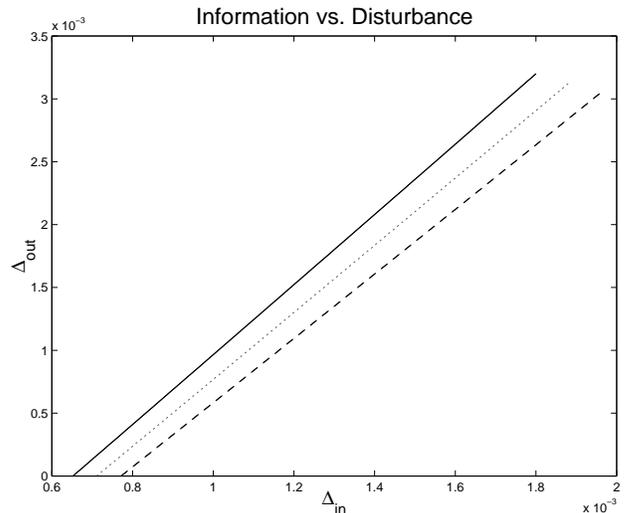,width=3.25in}}
\caption{\narrowtext The tradeoff between information, 
$\Delta_{\ho in}$, and disturbance, $\Delta_{\ho out}$, is 
plotted here for $\alpha=1$, $b = 0.1$ and three values of 
$a$. Solid line: $a=0.8$, dotted line: $a=0.79$ and 
dot-dash line: $a=0.78$ .}
\label{fig3}
\end{figure}

The general tradeoff relation, when it exists, is found simply
enough by eliminating the variable $z$ from the simultaneous
equations (\ref{JoelUnderwood}) and (\ref{DaleUnderwood}). (Two 
examples of the tradeoff relation are given in figures \ref{fig2} 
and \ref{fig3}.) This
time---in contrast to what we did in Eq.~(\ref{NoDiscount})
however---we leave finding the explicit expression as an exercise
for the reader:  Seeing it explicitly adds little to the analysis
already given.

\section{Discussion}

Our conclusion is straightforward:  There are regions in the space of
finite-strength efficient measurements without feedback for which a
nontrivial information tradeoff relation exists as one unitarily
varies around any given fiducial measurement $\cal M$. In a way, it
is a shame that we could not make a more unqualified assertion---for
instance, that a nontrivial tradeoff relation held for {\it all\/}
finite strength quantum measurements without feedback.  Indeed the
hope that such would be the case was a large part of the motivation
for this work.

The question now arises as to the significance of the rather
complicated regions defined by Eqs.~(\ref{Django}) and
(\ref{Reinhardt}).  What trenchant physical property is implied of a
measurement $\cal M$ that sits in the information-disturbance region
of a given density operator $\rho$?

A toy idea is that the key distinction lies not in finite vs.\
infinite measurement strength, but in whether the measurement sits
above or below a certain finite-strength 

$\!\!\!\!\!\!$ threshold. That is to say,
in carrying out the program of this paper, we would imagine not only
varying over unitary orbits for defining a tradeoff relation, but
rather over any region of POVM space so long as a certain constraint
on the measurement strength is obeyed. Unfortunately, if this is
going to be the case, it is going to require some thinking more
subtle than we have carried out so far.  This is because for at least
one natural definition of measurement strength we again find no
nontrivial tradeoff relation. The failure of this program is
described in Appendix B\@. So the question remains.

In general, this paper forms part of a larger effort to fully
delimit the information-disturbance tradeoff properties of quantum
mechanics.

\section*{Appendix A: Efficient Measurements Increase Alice's
Information}

In this appendix, we prove that for any efficient quantum
measurement, an observer aware of the outcomes will on average
increase his ``knowledge'' of the measured quantum system. More
precisely, when a measurement causes the observer to update his
density operator from $\rho$ to $\rho_b$---as in
Eq.~(\ref{Aharonov})---it holds for any concave unitarily invariant
function $F$ that
\be
F(\rho)\ge\sum_b p_b F(\rho_b)\;.
\label{ShakeyPie}
\ee
Along the way, and as something of an aside, we will also prove a
stronger result that deals directly with relations between the
eigenvalues of $\rho$ and all the $\rho_b$.  This result is most
conveniently couched in terms of the mathematical theory of
majorization \cite{Marshall79}, and will require a little notation
for its statement.

Let us define $\vec{\lambda}({\cal O})$ to be the vector of
eigenvalues of a Hermitian operator $\cal O$ on ${\cal H}_d$, with
the components arranged in terms of decreasing magnitude. That is to
say, let the numbers $\lambda_i({\cal O})$ obey the ordering
\be
\lambda_1({\cal O})\ge\lambda_2({\cal O})\ge\cdots\ge\lambda_d({\cal
O})\;.
\ee
We say that a vector $\vec{\lambda}({\cal O})$ is {\it majorized\/}
by a vector $\vec{\lambda}({\cal N})$, and write
\be
\vec{\lambda}({\cal O})\prec\vec{\lambda}({\cal N})
\ee
when
\be
\sum_{i=1}^k \lambda_i({\cal O})\le \sum_{i=1}^k \lambda_i({\cal N})
\ee
for all $k=1,2,\ldots,d$, and
\be
\sum_{i=1}^d \lambda_i({\cal O})= \sum_{i=1}^d \lambda_i({\cal N})\;.
\ee
One can also say that a Hermitian operator $\cal O$ is {\it
majorized\/} by a Hermitian operator $\cal N$, and write ${\cal
O}\prec{\cal N}$, when $\vec{\lambda}({\cal
O})\prec\vec{\lambda}({\cal N})$, but we will not have any need for
that terminology in this development.

Our main result is this:
\be
\vec{\lambda}(\rho)\prec\sum_b p_b \vec{\lambda}(\rho_b)
\label{Schninket}
\ee
(As pointed out before, this result has also been obtained recently
in Ref.~\cite{Nielsen00}.) The proof of this statement is not too
difficult if we rely on some results from the mathematical literature
\cite{Marshall79}, and a method of thought promoted by Ben Schumacher
on several occasions to great result---see
Refs.~\cite{Schumacher96,Linden00}, to name only a few.

The trick of Schumacher is this.  Whenever we have a quantum system
$Q$ and we say that it is in a (mixed) state $\rho$, there is
nothing to prevent us from thinking that the situation has come about
because $Q$ is part of a larger system $RQ$, which we happen to
describe via some pure state $|\psi^{\ho RQ}\rangle$.  The state
$\rho$ then is just a partial trace over the larger pure state:
\be
\rho={\rm tr}_{\ho R}|\psi^{\ho RQ}\rangle\langle\psi^{\ho RQ}|
\ee
There are times when such a conception can be quite useful for
simplifying the mathematics of a problem.  The problem at hand is one
of them.

Let us now describe the measurement process on a system $Q$ from such
a point of view.  It will be useful to make explicit precisely which
system we are referring to at any given time:  therefore we shall add
superscripts or subscripts, $R$, $Q$, or $RQ$ to all density
operators to make that clear.  In these new terms, the state change
under measurement that we are interested in is given by
\be
\rho^{\ho Q}\;\longrightarrow\;\rho^{\ho Q}_b=\frac{1}{p_b}U_b
E_b^{1/2}\rho^{\ho Q}E_b^{1/2}U_b^\dagger
\ee
after an outcome $b$ is found.  That same measurement, on the other
hand, changes the state of the $RQ$ system according to:
\be
|\psi^{\ho RQ}\rangle\;\longrightarrow\; |\psi^{\ho RQ}_b\rangle =
\sqrt{\frac{1}{p_b}}\,\big(I_{\ho R}\!\otimes A_b\big) |\psi^{\ho
RQ}\rangle\;,
\ee
where
\be
A_b = U_b E_b^{1/2}
\ee 
(Recall that pure states remain pure under an efficient
measurement.)  The operator $I_{\ho R}$ in this equation, of course,
signifies the identity operator on the $R$ system.

Note that the initial density operators $\rho^{\ho R}$ and $\rho^{\ho
Q}$ for the $R$ and $Q$ systems are unitarily equivalent.  I.e.,
\be
\rho^{\ho R} = {\rm tr}_{\ho Q}|\psi^{\ho RQ}\rangle\langle\psi^{\ho
RQ}|=V \rho^Q V^\dagger\;,
\ee
for some unitary operator $V$.  In particular, it follows that
$\rho^{\ho R}$ and $\rho^{\ho Q}$ have the same eigenvalues.  We can
also note, however, that since a measurement on $Q$ can have no
overall effect on the $R$, it must be the case that
\be
\rho^{\ho R}={\rm tr}_{\ho Q}\!\left(\sum_b p_b|\psi^{\ho
RQ}_b\rangle\langle\psi^{\ho RQ}_b|\right).
\label{Altoid}
\ee
One can see this more formally by choosing a Schmidt decomposition
for $|\psi^{\ho RQ}\rangle$,
\be
|\psi^{\ho RQ}\rangle = \sum_{k=1}^d
\sqrt{\lambda_k}|r_k\rangle|q_k\rangle \;.
\ee
Then
\beq
{\rm tr}_{\ho Q}&&\!\left(\sum_b p_b|\psi^{\ho
RQ}_b\rangle\langle\psi^{\ho RQ}_b|\right) =
\\
&&=\sum_{lb}\langle q_l| \big(I_{\ho R}\!\otimes A_b\big) |\psi^{\ho
RQ}\rangle\langle\psi^{\ho RQ}|\big(I_{\ho R}\!\otimes
A_b^\dagger\big)|q_l\rangle
\\
&&=\sum_{kmlb}\sqrt{\lambda_k}\sqrt{\lambda_m}|r_k\rangle\langle
r_m|\langle q_l|A_b|q_k\rangle \langle q_m|A_b^\dagger|q_l\rangle
\\
&&=\sum_{kmlb}\sqrt{\lambda_k}\sqrt{\lambda_m}|r_k\rangle\langle
r_m|\langle q_m|A_b^\dagger|q_l\rangle \langle q_l|A_b|q_k\rangle
\\
&&=\sum_{kmb}\sqrt{\lambda_k}\sqrt{\lambda_m}|r_k\rangle\langle
r_m|\langle q_m|A_b^\dagger A_b|q_k\rangle
\\
&&=\sum_{km}\sqrt{\lambda_k}\sqrt{\lambda_m}|r_k\rangle\langle
r_m|\delta_{mk}
\\
&&=\sum_{k}\lambda_k|r_k\rangle\langle r_k|
\\
&&=\rho^{\ho R}\;.
\eeq

It follows from Eq.~(\ref{Altoid}) and the statement preceding it
that
\be
\vec{\lambda}(\rho^{\ho Q})=\vec{\lambda}(\rho^{\ho R})=
\vec{\lambda}\left({\rm tr}_{\ho Q}\!\left(\sum_b p_b|\psi^{\ho
RQ}_b\rangle\langle\psi^{\ho RQ}_b|\right)\right).
\ee

But ${\rm tr}_{\ho Q}$ is a linear mapping.  So defining
\be
\rho^{\ho R}_b = {\rm tr}_{\ho Q}|\psi^{\ho
RQ}_b\rangle\langle\psi^{\ho RQ}_b|\;,
\ee
we have
\be
\vec{\lambda}(\rho^{\ho Q})=\vec{\lambda}\left(\sum_b p_b\rho^{\ho
R}_b \right)\;.
\label{MinxKitten}
\ee

Now comes the point where we rely on the mathematical literature
ever so slightly by using Ky Fan's dominance theorem
\cite{Marshall79}. One can show that for any Hermitian operator $\cal
O$,
\be
\sum_{i=1}^k\lambda_i({\cal O})= \max_P\, {\rm tr}P{\cal O}\;,
\ee
where the maximization is taken over all rank-$k$ projectors. It
follows from this almost immediately that
\be
\vec{\lambda}({\cal O+N})\prec\vec{\lambda}({\cal
O})+\vec{\lambda}({\cal N})\;
\ee
since
\be
\max_P\, {\rm tr}P({\cal O+N})\le\max_P\, {\rm tr}P{\cal O}+\max_P\,
{\rm tr}P{\cal N}\;.
\ee
It follows from this that
\be
\vec{\lambda}(\rho^{\ho Q})\prec \sum_b \vec{\lambda}(p_b \rho^{\ho
R}_b)=\sum_b p_b\vec{\lambda}(\rho^{\ho R}_b)\;,
\ee
since $\vec{\lambda}(c{\cal O})=c\vec{\lambda}({\cal O})$ for any
positive number $c$.

Noting finally that the eigenvalue spectrum of $\rho^{\ho R}_b$ is
the same as that of $\rho^{\ho Q}_b$, we have ultimately
\be
\vec{\lambda}(\rho^{\ho Q})\prec \sum_b p_b\vec{\lambda}(\rho^{\ho
Q}_b)\;.
\ee
Stripping off the superscript $Q$, we have the desired result
Eq.~(\ref{Schninket}), and the theorem is proved.

It comes about as a corollary to Eq.~(\ref{Schninket}), through some
theorems in Ref.~\cite{Ando89} that our most desired result---namely
Eq.~(\ref{ShakeyPie})---holds for any concave unitarily invariant
function $F$. However, there is a more direct way to see this, and it
seems worthwhile to take that route. We need only back up to
Eq.~(\ref{MinxKitten}).  From this it follows that
\be
F(\rho^{\ho Q})=F\left(\sum_b p_b\rho^{\ho R}_b \right)\;.
\ee
But, $F$ is concave and so
\beq
F(\rho^{\ho Q}) &\ge& \sum_b p_b F(\rho^{\ho R}_b)
\\
&=& \sum_b p_b F(\rho^{\ho Q}_b)\;.
\eeq
Again, stripping off the superscript $Q$, we obtain the desired
result.

{\it Note Added in Proof:}  Instructive though it is to derive
Eq.~(\ref{Schninket}) by first extending the problem to an ancillary
Hilbert space, there is an even shorter route to the result that is
worth recording.  The trick to note is this: With each efficient
measurement ${\cal E}=(E_b)=(A_b^\dagger A_b)$, we can associate a
canonical decomposition of the density operator starting from the
fact that the $E_b$ form a resolution of the identity. Starting from
the equation
\be
I=\sum_b E_b\;,
\ee
one simply multiplies it from the left and right by $\rho^{1/2}$ to
get
\be
\rho=\sum_b p_b \omega_b
\label{OmegaMan}
\ee
where
\be
\omega_b=\frac{1}{p_b}\rho^{1/2}E_b \rho^{1/2}
\ee
and $p_b={\rm tr}\rho E_b$ as always.

Using the Ky Fan dominance theorem just as before, but now on
Eq.~(\ref{OmegaMan}), we have straight away that
\be
\vec{\lambda}(\rho)\prec \sum_b p_b \vec{\lambda}(\omega_b)\;.
\ee
However, it is an easy matter to see that the operators $\rho^{1/2}
E_b \rho^{1/2}$ and $A_b\rho A_b^\dagger$ have precisely the same
eigenvalue structure.  Just start off with the eigenvalue equation
\be
\big(\rho^{1/2}A_b^\dagger
A_b\rho^{1/2}\big)|i\rangle=\mu_i|i\rangle\;.
\ee
Multiplying this from the left by $A_b\rho^{1/2}$ and regrouping
terms, one gets
\be
A_b\rho A_b^\dagger \big(A_b\rho^{1/2}|i\rangle\big)=\mu_i
\big(A_b\rho^{1/2}|i\rangle\big)\;,
\ee
which means that $\rho^{1/2} E_b \rho^{1/2}$ and $A_b\rho
A_b^\dagger$ have the same eigenvalues.  Using this,
Eq.~(\ref{Schninket}) follows immediately.

\section*{Appendix B:  Full Variation over Measurements of a Given
Strength}

For this Appendix, we drop the distinction of finite vs.\ infinite
measurement strength and attempt to grade all measurements via a
single {\it finite\/} number.   One possible notion of such a
measurement strength is the amount by which Alice's purity would
change if $\rho$ happened to be the maximally mixed state
$\frac{1}{2}I$---that is, the measurement strength would be her
change of knowledge if she starts out completely ignorant of the
system. We can do this with respect to any of the functions $F$ in
Eq.~(\ref{Busch}), but for convenience we will again adopt the
impurity $P$ to be the main function of interest. Also for
convenience, we will actually adopt two times the said quantity
above, i.e., $2\Delta_{\ho in}(\frac{1}{2}I)$. This choice of
pre-factor causes our notion of measurement strength to range in the
full interval $[0,1]$.

Thus, using Eq.~(\ref{JoelUnderwood}) and taking $a=0$, a given
measurement strength $k$ for a two-outcome measurement $(E,I-E)$ is
defined by
\be
k=\frac{\alpha b^2}{2-\alpha}\;.
\label{Rufus}
\ee
The question we shall pose in this Appendix is the following.  For a
given quantum state $\rho$ and a fixed measurement strength $k$,
what is the maximum value of $\Delta_{\ho in}$ and what values of $z$
achieve that maximum?  In particular, can we show that the optimal
values for $z^2$ in this problem are strictly less than 1?
Unfortunately, we will have to answer the latter question in the
negative, regardless of the value $a$ defining the purity of the
initial density operator.

This is seen as follows.  Fix $k$ anywhere in the range between 0
and 1.  For a fixed $b$ this means that $\alpha$ must take on the
value
\be
\alpha=\frac{2k}{b^2+k}\;.
\label{Hodaddy}
\ee
Note that for a fixed value of $k$ we are not allowed to freely
choose $b$ as we wish.  This is because for a fixed $k$, the
variable $b$ cannot be too small or we would never be able to satisfy
Eq.~(\ref{Rufus}).  The valid range for $b$ turns out to be
\be
k\le b\le 1\;.
\label{Range-ola}
\ee
The consideration leading to this is simple.  The function
$\frac{\alpha}{2-\alpha}$ is monotonically increasing in $\alpha$.
So to find our smallest value of $b$, we should place the largest
allowed value of $\alpha$, Eq.~(\ref{Mojo}), into the right-hand side
of Eq.~(\ref{Hodaddy}). Doing this gives Eq.~(\ref{Range-ola}).

Plugging Eq.~(\ref{Hodaddy}) into Eq.~(\ref{JoelUnderwood}) gives a
surprisingly simple expression
\be
\Delta_{\ho in}=\frac{k(1-a^2)(1-a^2z^2)b}{2(1+azb)(b-akz)}\;.
\ee
Let us now examine the behavior of this as a function of $b$. Taking
the partial derivative with respect to $b$, we obtain
\be
\frac{2}{k(1-a^2)(1-a^2 z^2)}\,\frac{\partial\Delta_{\ho
in}}{\partial b}\,=\,-\frac{a(b^2+k)z}{(1+azb)^2(b-akz)^2}\;.
\ee
Therefore $\Delta_{\ho in}(b)$ toggles from being an increasing to a
decreasing function depending upon the sign of $z$.  Thus $\max_b
\Delta_{\ho in}(b)$ takes on a piecewise form.  If $z\ge0$, we
should choose $b=k$; if $z\le0$, we should choose $b=1$.  The
resultant of these choices is conveniently summarized as follows:
\be
\Delta_{\ho in}^{\!\ho
max}(z)=\frac{1}{2}k(1-a^2)\frac{1+a|z|}{1+ak|z|}\;.
\label{MapleSyrple}
\ee

The function $\Delta_{\ho in}^{\!\ho max}(z)$ in
Eq.~(\ref{MapleSyrple}) is increasing in $|z|$ since $k\le1$.  Hence
it finally follows that the very best strategy on Alice's part for a
given measurement strength $k$ is to take $z=1$ or $-1$.  Doing so
gives her an absolute maximum purity change of
\be
\Delta_{\ho in}^{\!\ho max}=\frac{1}{2}k(1-a^2)\frac{1+a}{1+ak}\;,
\ee
and that purity change is accompanied by a purity change of
$\Delta_{\ho out}=0$ for Bob.

\section*{Acknowledgements}

We thank Tanmoy Bhattacharrya, Salman Habib, Alexander Holevo, and
Ben Schumacher for helpful discussions.
%This research was performed
%in part using resources located at the Advanced Computing Laboratory
%at Los Alamos National Laboratory.


\begin{references}
\bibitem{EPR}
A.~Einstein, B.~Podolsky, and N.~Rosen, Phys.\ Rev.\ {\bf 47},
777--780 (1935).

\bibitem{Bohr35}
N.~Bohr, Phys.\ Rev.\ {\bf 48}, 696--702 (1935).

\bibitem{CommentNonsense}
We should point out that lately it has become fashionable to say
that Bohr dropped his rhetoric of ``measurement causing
disturbance'' soon after his reply to the EPR paper
\cite{MerminPrivate}. We disagree with this as the counter-evidence
is easily exhibited in almost everything that Bohr wrote on the
subject thereafter.  A prime example is this passage from
Ref.~\cite{Bohr48}.

\begin{quotation}
The very fact that quantum phenomena cannot be analysed on classical
lines thus implies the impossibility of separating a behavior of
atomic objects from the interaction of these objects with the
measuring instruments which serve to specify the conditions under
which the phenomena appear.  In particular, the individuality of the
typical quantum effects finds proper expression in the circumstance
that any attempt at subdividing the phenomena will demand a change
in the experimental arrangement, introducing new sources of
uncontrollable interaction between objects and measuring instruments
\end{quotation}

\bibitem{MerminPrivate}
N.~D. Mermin, private communication, 19 May 2000.

\bibitem{Bohr48}
N.~Bohr, Dialectica {\bf 2}, 312--318 (1948).

\bibitem{Pauli95}
W.~Pauli, {\sl Writings on Philosophy and Physics}, edited by C.~P.
Enz and K. von Meyenn, (Springer-Verlag, Berlin, 1995), p.~132.

\bibitem{Jammer74}
M.~Jammer, {\sl  The Philosophy of Quantum Mechanics: The
Interpretations of Quantum Mechanics in Historical Perspective},
(Wiley, New York, 1974).

\bibitem{Beller99}
M.~Beller, {\sl Quantum Dialogue: The Making of a Revolution}, (U. of
Chicago Press, Chicago, 1999).

\bibitem{NoBohm}
We pay no attention to {\it variants\/} of quantum mechanics, such as
Bohmian mechanics, that incorporate nonlocal hidden variables.

\bibitem{Mermin90}
For a clear discussion of this point, see N.~D. Mermin, {\sl Boojums
All the Way Through:\ Communicating Science in a Prosaic Age},
(Cambridge U. Press, Cambridge, 1990), pp.~110--176.

\bibitem{Peres00}
A.~Peres, Phys.\ Rev.\ A {\bf 61}, 022116-1--022116-9 (2000).

\bibitem{Fuchs00}
C.~A. Fuchs and A.~Peres, Phys.\ Today {\bf 53}(3), 70--71 (2000).

\bibitem{Braunstein88}
S.~L. Braunstein and C.~M. Caves, Found.\ Phys.\ Lett.\ {\bf 1},
3--12 (1988).

\bibitem{WheelerZurek}
W.~Heisenberg, in {\sl Quantum Theory and Measurement}, edited by
J.~A. Wheeler and W.~H. Zurek (Princeton U. Press, Princeton, NJ,
1983), pp.~62--84.

\bibitem{Unruh78}
W.~G. Unruh, Phys.\ Rev.\ D {\bf 18}, 1764--1772 (1978); {\bf 19},
2888--2896 (1979).

\bibitem{Kraus83}
K.~Kraus, {\sl States, Effects, and Operations:\ Fundamental Notions
of Quantum Theory}, (Springer-Verlag, Berlin, 1983).

\bibitem{Wiesner83}
The first appearance of this idea seems to be in the work of Stephen
Wiesner circa 1970, but remained unpublished until much later.  See
S.~Wiesner, SIGACT News {\bf 15}, 78--88 (1983).

\bibitem{Bennett84}
C.~H. Bennett and G.~Brassard, in {\sl Proc.\ IEEE International
Conference on Computers, Systems and Signal Processing, Bangalore,
India, December 10-12, 1984}, (IEEE Press, New York, 1984),
pp.~175--179.

\bibitem{Bennett92a}
C.~H. Bennett, G.~Brassard, and N.~D. Mermin, Phys.\ Rev.\ Lett.\
{\bf 68}, 557--559 (1992).

\bibitem{Bennett92b}
C.~H. Bennett, Phys.\ Rev.\ Lett.\ {\bf 68}, 3121--3124 (1992).

\bibitem{Fuchs98}
C.~A. Fuchs, Fort.\ der Phys.\ {\bf 46}, 535--565 (1998).

\bibitem{Fuchs96}
C.~A. Fuchs and A.~Peres, Phys.\ Rev.\ A {\bf 53}, 2038--2045 (1996).

\bibitem{Fuchs97}
C.~A. Fuchs, N.~Gisin, R.~B. Griffiths, C.-S. Niu, and A.~Peres,
Phys.\ Rev.\ A {\bf 56},  1163--1172 (1997).

\bibitem{Bruss98}
D.~Bruss, Phys.\ Rev.\ Lett.\ {\bf 81}, 2598--2601 (1998).

\bibitem{Kyburg80}
H.~E. Kyburg, Jr.\ and H.~E. Smokler, eds., {\sl Studies in
Subjective Probability}, Second Edition, (Robert~E. Krieger
Publishing, Huntington, NY, 1980).

\bibitem{Bernardo94}
J.~M. Bernardo and A.~F.~M. Smith, {\sl Bayesian Theory}, (Wiley, New
York, 1994).

\bibitem{FuchsInfinity}
One of the authors (CAF) is tempted to speculate that this idea alone
(suitably refined) captures the essence of quantum mechanics. The
formalism of quantum theory, by this view, is simply the best
agreement we can all come to in a world so sensitive to our
experimental interventions. This contrasts with the defining property
of the classical world, which, at the outset, is assumed describable
by a set of variables stable enough that we can discover them or, at
the very least, safely contemplate their supposed existence.

\bibitem{Gleason57}
A.~M. Gleason, J. Math.\ Mech.\ {\bf 6}, 885--894 (1957).

\bibitem{Ash65}
R.~B. Ash, {\sl Information Theory}, (Dover, New York, 1965).

\bibitem{Wehrl78}
A.~Wehrl, Rev.\ Mod.\ Phys.\ {\bf 50}, 221--259 (1978).

\bibitem{Wootters89}
W.~K. Wootters and B.~D. Fields, Ann.\ Phys.\ {\bf 191}, 363--381
(1989).

\bibitem{Wootters90}
W.~K. Wootters, Found.\ Phys.\ {\bf 20}, 1365--1378 (1990).

\bibitem{Jones91}
K.~R.~W. Jones, J. Phys.\ A {\bf 24}, 1237--1244 (1991).

\bibitem{Luders51}
G.~L\"uders, Ann.\ der Phys.\ {\bf 8}, 323--328 (1951).

\bibitem{Busch96}
P.~Busch, P.~Lahti, and P.~Mittelstaedt, {\sl The Quantum Theory of
Measurement\/}, Second revised edition, (Springer-Verlag, Berlin,
1996).

\bibitem{Busch98}
P.~Busch and J.~Singh, Phys.\ Lett.\ A {\bf 249}, 10--12 (1998).

\bibitem{Nielsen00}
M.~A. Nielsen, ``Characterizing Mixing and Measurement in Quantum
Mechanics,'' {\tt quant-ph/0008073}.

\bibitem{Peres93}
A.~Peres, {\sl Quantum Theory: Concepts and Methods}, (Kluwer,
Dordrecht, 1993).

\bibitem{WM} H.M. Wiseman and G.J. Milburn, Phys. Rev. A {\bf 47}, 
642 (1993).

\bibitem{Schatten60}
R.~Schatten, {\em Norm Ideals of Completely Continuous Operators},
(Springer-Verlag, Berlin, 1960).

\bibitem{Barnum98}
H.~N. Barnum, {\sl Quantum Information Theory}, Ph.\ D. Thesis,
University of New Mexico, Albuquerque, NM, 1998, Chap.~4.

\bibitem{Banaszek00}
K.~Banaszek, ``Fidelity Tradeoff in Quantum Operations,'' LANL
e-print archive, {\tt quant-ph/0003123}.

\bibitem{Doherty00}
A.~C. Doherty, K.~Jacobs, and G.~Jungman, ``Information, Disturbance,
and Hamiltonian Quantum Feedback Control,'' {\tt quant-ph/0006013}.

\bibitem{Fujiwara98}
A.~Fujiwara and H.~Nagaoka, IEEE Trans.\ Inf.\ Theory {\bf 44},
1071--1086 (1998).

\bibitem{Gallager68}
R.~G. Gallager, {\sl Information Theory and Reliable Communication},
(Wiley, New York, 1968).

\bibitem{Jozsa94}
R.~Jozsa, D.~Robb, and W.~K. Wootters, Phys.\ Rev.\ A {\bf 49},
668-677 (1994).

\bibitem{Caves96}
C.~M. Caves and C.~A. Fuchs, in {\sl The Dilemma of Einstein,
Podolsky and Rosen -- 60 Years Later}, Ann. Israel Phys.\ Soc.\ {\bf
12}, edited by A.~Mann and M.~Revzen, 226--257 (1996).

\bibitem{Jaynes57}
E.~T. Jaynes, Phys.\ Rev. {\bf 106}, 620--630 (1957).

\bibitem{Aczel75}
J. Acz\'{e}l and Z. Dar\'{o}czy, {\sl On Measures of Information and
Their Characterizations}, (Academic Press, New York, 1975).

\bibitem{Nielsen00b}
M.~A. Nielsen also makes a similar point in Ref.~\cite{Nielsen00}.

\bibitem{Ando89}
T.~Ando, Lin.\ Alg.\ App.\ {\bf 118}, 163--248 (1989).

\bibitem{Marshall79}
A.~W. Marshall and I.~Olkin, {\sl Inequalities:~Theory of
Majorization and Its Applications}, (Academic Press, New York, 1979).

\bibitem{Schumacher96}
B.~W. Schumacher, Phys.\ Rev.\ A {\bf 54}, 2614--2628 (1996).

\bibitem{Linden00}
N.~Linden, S.~Popescu, B.~Schumacher, and M.~Westmoreland,
``Reversibility of Local Transformations of Multiparticle
Entanglement,'' {\tt quant-ph/9912039}.

\end{references}
\end{document}